\documentclass[
reprint,
preprintnumbers,
nofootinbib,
amsmath,amssymb,longbibliography,
aps,
prd
]{revtex4-1}
\pdfoutput=1
\usepackage{graphicx}
\usepackage[utf8]{inputenc}
\usepackage{flushend}
\usepackage{dcolumn}
\usepackage{bm}
\usepackage{soul}
\usepackage{balance}
\usepackage[normalem]{ulem}
\usepackage[colorlinks = true,
            linkcolor = blue,
            urlcolor  = blue,
            citecolor = blue,
            anchorcolor = blue]{hyperref}
\usepackage{verbatim,subfigure}
\usepackage{color,ulem}
\usepackage[english]{babel}
\usepackage{MnSymbol,wasysym}
\usepackage[utf8]{inputenc}
\input Starburst.fd
\newcommand*\initfamily{\usefont{U}{Starburst}{xl}{n}}\initfamily 

\newcommand{\beq}{\begin{eqnarray}}
\newcommand{\eeq}{\end{eqnarray}}
\usepackage{amsmath}
\usepackage{tikz}
\usetikzlibrary{decorations.pathmorphing}
\usetikzlibrary{shapes.misc}
\tikzset{cross/.style={cross out, draw=black, minimum size=8*(#1-\pgflinewidth), inner sep=0pt, outer sep=0pt},
cross/.default={1pt}}
\usetikzlibrary{patterns,math}
\definecolor{applegreen}{rgb}{0.55, 0.71, 0.0}
\usepackage{contour}
\usepackage{xparse}
\ExplSyntaxOn
\NewDocumentCommand{\HS}{m}
 {
  \seq_set_split:Nnn \l_tmpa_seq { ~ } { #1 }
  \seq_map_inline:Nn \l_tmpa_seq { \contour{green}{##1} ~ } \unskip
 }
\ExplSyntaxOff

\definecolor{darkviolet}{rgb}{0.58, 0.0, 0.83}
\definecolor{mygreen}{rgb}{0.0, 0.5, 0.0}

\begin{document}

%
\title{Tracking shear mode dynamics across the glass transition in a 2D colloidal system}
\author{Jimin Bai$^{1}$}
\author{Peter Keim$^{2,3,4}$}
\email{peter.keim@uni-duesseldorf.de}
\author{Matteo Baggioli$^{1}$}
\email{b.matteo@sjtu.edu.cn}

\affiliation{$^1$Wilczek Quantum Center, School of Physics and Astronomy, Shanghai Jiao Tong University, Shanghai 200240, China \& Shanghai Research Center for Quantum Sciences, Shanghai 201315, China}
\affiliation{$^2$Institut for Experimental Physics of Condensed Matter, Heinrich-Heine-Universit\"at D\"usseldorf, 40225 D\"usseldorf, Germany}
\affiliation{$^3$Max-Planck-Institute for Dynamics and Self-Organization, 37077 G\"ottingen, Germany}
\affiliation{$^4$Institute for the Dynamics of Complex Systems, University of G\"ottingen, 37077 G\"ottingen, Germany}

\begin{abstract}
Long-wavelength collective shear dynamics are profoundly different in solids and liquids. According to the theoretical framework developed by Maxwell and Frenkel, collective shear waves vanish upon melting by acquiring a characteristic wave-vector gap, known as the $k$-gap. While this prediction has been supported by numerous simulations, experimental validation remains limited. Moreover, this phenomenon has been never tested across a continuous glass transition between a liquid phase and a glassy state with large but finite viscosity. In this work, we track the dispersion relation of collective shear modes in a two-dimensional colloidal system and provide direct experimental evidence for the emergence of a $k$-gap. This gap opens continuously at an effective temperature consistent with the onset of the glass transition and the vanishing of the static shear modulus. By extracting the instantaneous shear velocity from the experimental data, we uncover a shear relaxation time exhibiting a super-Arrhenius temperature dependence characteristic of glass-forming materials, accurately described by the Vogel–Fulcher–Tammann (VFT) relation. Our results confirm the predictions of the Maxwell-Frenkel framework and highlight their relevance across continuous melting processes originating from low-temperature amorphous solid phases.
\end{abstract}

\maketitle
%
\section{Introduction}
At low wave vectors, collective shear dynamics in solids are governed by propagating shear waves (phonons), with their speed determined by the static shear modulus \cite{chaikin1995principles}. In contrast, in liquids, these dynamics are dominated by diffusive processes governed by the finite shear viscosity \cite{landau2013fluid}. 

Within the theoretical frameworks established by Maxwell and Frenkel \cite{Trachenko_2016}, the transition between these two regimes is governed by the telegrapher equation:
\begin{equation}
    \omega_T^2+i \omega_T/\tau = v^2 k^2,\label{eq1}
\end{equation}
where $\omega_T$ is the frequency of collective transverse excitations, $k$ the wave vector, $v$ the instantaneous speed of sound and $\tau$ a characteristic relaxation time. In Frenkel’s microscopic picture of liquid dynamics \cite{Frenkel1946}, $\tau$ represents the average time a particle takes to hop over potential barriers, that can be related to the lifetime of local atomic connectivity \cite{huang2025atomisticmechanismsviscosity2d}. In contrast, Maxwell’s theory \cite{maxwell1867iv} identifies $\tau$ as the Maxwell relaxation time $\tau_M$, which characterizes the macroscopic viscoelastic response of the medium and is formally defined as the ratio between the shear viscosity $\eta$ and the instantaneous shear modulus $G_{\infty}$, i.e., $\tau_M=\eta/G_{\infty}$.

Equation~\eqref{eq1}, which arises in a wide range of physical systems \cite{BAGGIOLI20201}, predicts that collective shear excitations in liquids exhibit a characteristic dispersion:
\begin{equation}
    \mathrm{Re}(\omega_T)= v \sqrt{k^2-k_g^2},\label{eq2}
\end{equation}
where $k_g \equiv 1/(2 v \tau)$ is the so-called $k$-gap, characterizing the inverse length scale below which elastic response persists. This relation, Eq.~\eqref{eq2}, is well-established and can be formally derived using several methods, including generalized hydrodynamics (see chapter 6 in \cite{boon1991molecular}). In the long-wavelength limit, $k\ll k_g$, $\mathrm{Re}(\omega_T)=0$ and the dynamics are liquid-like (diffusive). In the other limit, $k \gg k_g$, solid-like response and propagating shear waves emerge also in liquids. This reflects the high-frequency shear modulus that makes diving from a $10$-meter platform so exciting, while swimming gently through water remains so relaxing.

\begin{figure*}
    \centering
    \includegraphics[width=\linewidth]{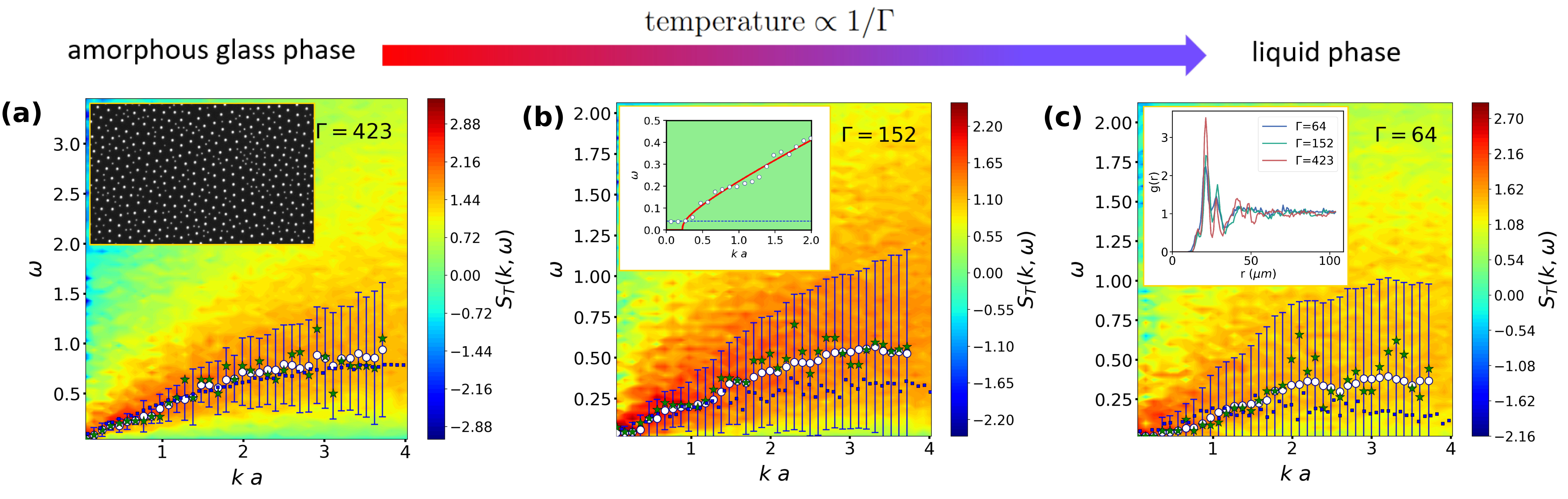}
    \caption{\textbf{(a)} The dispersion of collective shear modes as a function of the normalized wave-vector $k a$ for $\Gamma=423$, deep in the amorphous glass phase. White symbols are the value of $\Omega_T(k)$ obtained by fitting the dynamical transverse structure factor $S_T(k,\omega)$ with a damped harmonic oscillator (DHO) function and the error bars indicate the corresponding linewidth $\Gamma(k)$; green symbols are the position of the maxima in $S_T(k,\omega)$; blue symbols are the eigenvalues $\omega_j$ obtained diagonalizing the transverse sector of the dynamical matrix $\mathcal{D}(\mathbf{k})$. The background color is the absolute value of $S_T(k,\omega)$ in a 10-based logarithmic scale. The inset shows an image of the experimental setup. \textbf{(b)} Similar analysis for $\Gamma=152$, slightly below the expected glass transition temperature, in the liquid phase. The inset shows an example of the fit of the dispersion relation using the $k$-gap equation, Eq. \eqref{eq2}, with the horizontal dashed line indicating the cutoff frequency $\omega_{\text{min}}$ arising because of the finite size system. \textbf{(c)} Same analysis for $\Gamma=64$, deep in the liquid phase. The inset shows the value of the pair correlation function $g(r)$ for different values of $\Gamma$. The glass transition temperature for this system has been estimated to be around $\Gamma_g \approx 195$ \cite{PhysRevX.5.041033}.}
    \label{fig1}
\end{figure*}

Although the precise definition of $\tau$ in Eq.~\eqref{eq1} remains debated, it is generally expected that $k_g$ vanishes in crystalline solids, where the wave-like transverse phonon dispersion $\omega_T = v k$ dictated by elasticity theory is recovered. This follows directly from the fact that, in crystalline solids, $\tau \to \infty$, because the shear viscosity formally diverges. On the other hand, the behavior of $k_g$ across a continuous glass transition between a supercooled liquid and a glassy phase is more subtle. Since the viscosity in the glassy state does not diverge sharply, as it does across a first-order solidification transition, the corresponding $k_{\mathrm{gap}}$ is expected to approach zero continuously.

Since the initial validation of these theoretical predictions in simulated molecular liquids and supercritical fluids \cite{PhysRevLett.118.215502}, the $k$-gapped behavior described by Eq. \eqref{eq2} has been observed across a wide range of simulated systems, confirming its universality with respect to interparticle interactions (see \cite{Trachenko_2016} for a review).

On the other hand, experimentally verifying this mechanism in molecular liquids remains challenging due to limitations of current techniques in accessing the low-frequency, low-wave-vector regime. In \cite{PhysRevLett.97.115001}, the emergence of a cutoff wave number $k_g$ in the liquid-like phase of a two-dimensional Yukawa dusty plasma was observed and estimated as $k_g a \approx 0.16$–$0.31$, where $a$ is the 2D Wigner-Seitz radius. More recently, \cite{Jiang2025} reported the existence of a $k$-gap in the dispersion of collective transverse excitations in the liquid-like phase of athermal vibrated granular matter. However, in both cases, the disappearance of the wave-vector gap at the onset of solidification was not clearly demonstrated. 

It is worth noting that the concept of a phonon gap has been previously discussed both theoretically and experimentally in various contexts, including liquids, lipid membranes, and biological systems \cite{BOLMATOV20182446,Zhernenkov2016,Bolmatov2022}. However, whether this mechanism remains valid across a continuous glass transition from an amorphous solid to a liquid remains an open question that, to the best of our knowledge, has not yet been investigated. This is the main scope of our work.

We notice that, despite the full dispersion of shear waves across a glass transition has never been analyzed experimentally, the appearance of high-frequency shear waves across $T_g$ have been verified experimentally using Brillouin scattering \cite{PhysRevLett.62.2616}. The purpose of our analysis is to resolve the concrete dynamics behind this phenomenon and, thanks to the ability of probing low enough frequencies and wave-vectors, to probe the full dispersion relation of shear fluctuations across this continuous transition.

\section{Experimental system and data analysis}
Here, we study a two-dimensional colloidal glass composed of a binary mixture of super-paramagnetic polystyrene spheres confined to a flat water–air interface \cite{10.1063/1.3188948} (see the inset of Fig.~\ref{fig1}(a) for an image of the experimental setup). The small and large particles have diameters of $\sigma_S=2.8\,\mu m$ and $\sigma_L=4.5\,\mu m$, respectively, with a mixing ratio of $N_S:N_L=45:55$. Particle number is about $2300$ in the field of view, while the monolayer consists of about $100.000$ particles in total. The particles interact via a tunable magnetic dipole–dipole interaction, controlled by an external magnetic field and quantified by a dimensionless coupling parameter $\Gamma$, which serves as an effective inverse temperature. More precisely, our control parameter is given by
\begin{equation}
    \Gamma=\frac{\mu_0}{4 \pi}\,\frac{H^2 (\pi n)^{3/2}}{k_B T} \left(\xi \chi_S+(1-\xi)\chi_L\right)^2, \label{defgamma}
\end{equation}
where $H$ is the external magnetic field, $\mu_0$ the vacuum magnetic permeability, $T$ the temperature, $\xi$ the relative concentration of small  particles, $n$ the two-dimensional number density and $\chi_{S,L}$ the magnetic susceptibility of small and large particles respectively.

Particle positions are tracked over time using video microscopy and digital image analysis \cite{Crocker1996b}. This method, made possible by the mesoscopic size of our colloidal particles, overcomes the significant limitations of scattering techniques used in molecular liquids. These limitations are among the main reasons why the $k$-gap has never been convincingly demonstrated experimentally in those systems. The elastic properties of this experimental system have been thoroughly investigated \cite{PhysRevLett.109.178301}, with evidence suggesting the onset of a glass transition at $\Gamma_g \approx 195$ \cite{PhysRevX.5.041033}.

Following \cite{PhysRevLett.109.178301}, we define the particle displacement $\mathbf{u}_i(t) = \mathbf{r}_i(t)-\bar{\mathbf{r}}_i$, where $\bar{\mathbf{r}}_i$ is the average position of the particle $i \in [1,N]$ during a time interval $\Delta t \approx 18900$ s. Consistent with the analysis in \cite{PhysRevLett.109.178301}, the time interval is chosen to be sufficiently large such that the shear modulus vanishes in the liquid phase, thereby avoiding short-time elastic effects that are also present in liquids. We then derive the `dynamical matrix', mapped via the equipartition theorem to the squared displacements, $\mathcal{D}(\mathbf{k})=k_B T \langle \mathbf{u}^*_{\mathbf{k}}\mathbf{u}_{\mathbf{k}}\rangle ^{-1}$, where $\mathbf{u}_{\mathbf{k}}$ is the Fourier transform of the particle displacement in terms of the wave vector $\mathbf{k}$ and $\langle \cdot \rangle$ indicates time average. More specifically, the displacement field at each time frame in the dataset is calculated first, then the Fourier transform and Helmholtz decomposition is performed to separate the longitudinal and transverse dispersion, after which time average is performed over approximately $6000$ time frames to find the corresponding eigenvalues. 

The eigenvalues $\lambda_j(\mathbf{k})$ of the dynamical matrix correspond to the squared eigenfrequencies of the system $\omega_j^2(k)$ and provide the spectrum of excitations in the harmonic limit \cite{C2SM07445A}. Being overdamped by the solvent, the modes decay exponentially. The frequency $\omega_i$ of the modes are those of a ``shadow system'', where the solvent is absent but particle configuration and interactions remain unchanged \cite{C2SM07445A}.

From the mode analysis, we compute the transverse dynamical structure factor $S_T(k,\omega)$,
\begin{equation}
    S_{T}(k,\omega) \propto \frac{k^2}{\omega^2} \sum_\lambda E_{\lambda, T}(\mathbf{k})\delta \left(\omega-\omega_\lambda\right),
\end{equation}
where
\begin{equation}
    E_{\lambda, T}(\mathbf{k})=\Big| \sum_j \left(\hat{\mathbf{k}} \times \mathbf{e}_\lambda(j)\right)\exp\left(i \mathbf{k}\cdot \mathbf{r}_j\right)\Big|^2.
\end{equation}
Here, $\mathbf{r}_i$ is the position of the $i$th particle, $\mathbf{e}_\lambda(j)$ is the eigenvector corresponding to eigenfrequency $\omega_j$ and $\hat{\mathbf{k}}\equiv \mathbf{k}/|\mathbf{k}|$. To improve the statistics, we average $S_{T}(k,\omega)$ over different $\hat{\mathbf{k}}$ orientations.

\section{Results}
In Fig.~\ref{fig1}, we present the experimental dispersion relation of collective shear modes as a function of the dimensionless wave vector $k a$, where $a$ is the average interparticle distance $\approx 22\,\mu m$ (see inset of Fig. \ref{fig1}(c)), for several representative values of the control parameter $\Gamma$, spanning from deep within the amorphous solid phase (panel (a)) to the liquid phase (panel (c)). The background color map shows the absolute value of the transverse dynamical structure factor $S_T(k,\omega)$, plotted on a logarithmic scale to enhance contrast and visibility.

Blue filled symbols indicate the eigenvalues obtained by diagonalizing the dynamical matrix $\mathcal{D}(k)$. Green symbols mark the location of the maxima in $S_T(k,\omega)$ along constant-$k$ cuts. Finally, the white symbols represent the dispersion relation $\Omega_T(k)$ extracted from damped harmonic oscillator (DHO) fits to $S_T(k,\omega)$,
\begin{equation}
    S_T(k,\omega) \propto \frac{\omega^2 \Gamma_T(k)}{\left(\omega^2-\Omega^2_T(k)\right)^2+\omega^2 \Gamma_T(k)^2}.
\end{equation}
The error bars in Fig. \ref{fig1} indicate the relative linewidth $\Gamma_T(k)$.

We first observe that the three independent methods yield consistent results for the dispersion of collective shear modes in the solid phase. On the other hand, some deviations are observed in the liquid phase in the large wave-vector limit due to the the increased sound wave damping. More importantly, we note a progressive evolution in the form of the dispersion relation from panel (a), corresponding to $\Gamma = 423$, to panel (c), which corresponds to $\Gamma = 64$.

\begin{figure}[ht]
    \centering
    \includegraphics[width=\linewidth]{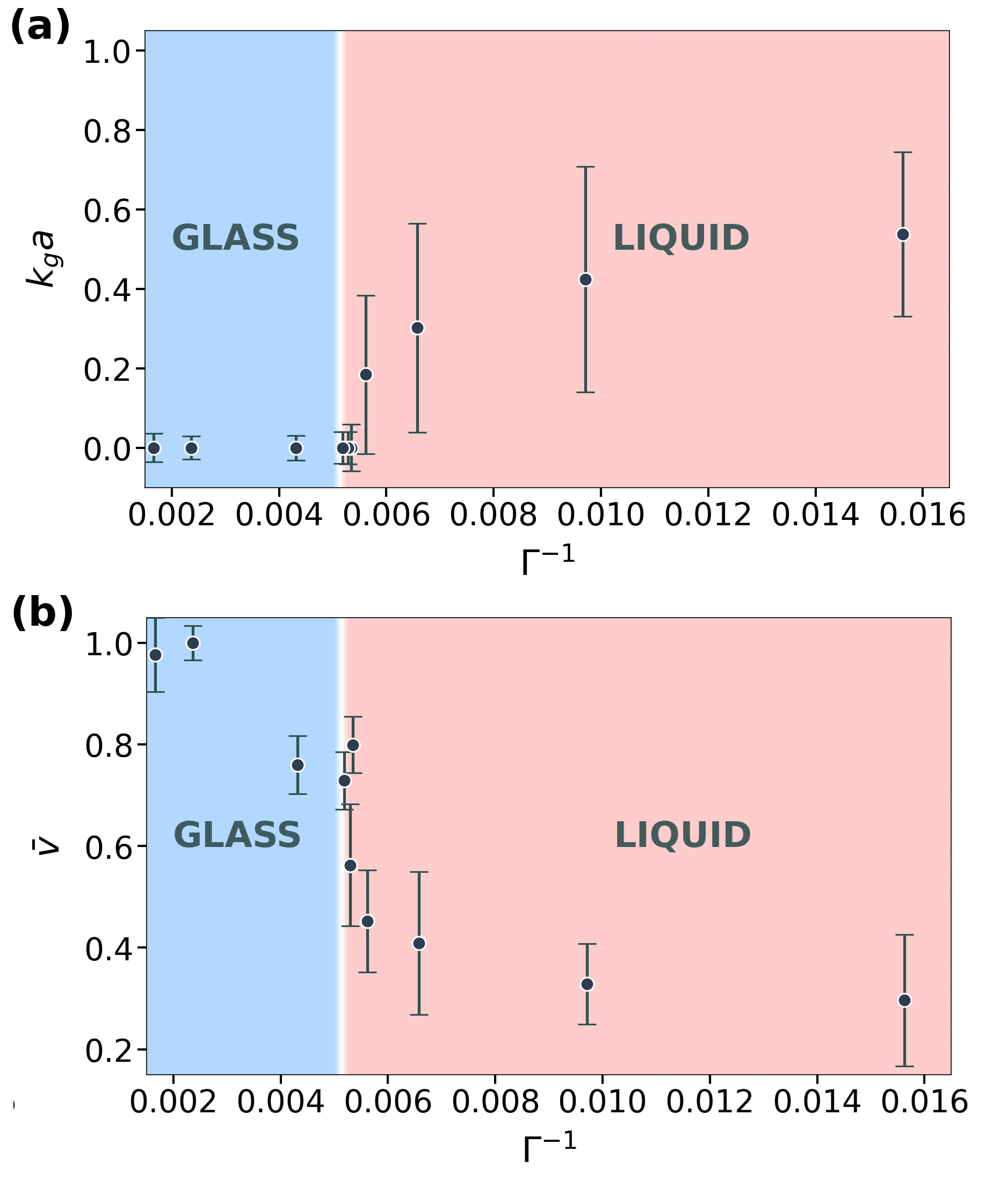}
    \caption{\textbf{(a)} The propagation gap for collective shear excitations $k_g$ in units of the Wigner-Seitz radius as a function of the inverse $\Gamma$ parameter, proportional to the temperature of the system. In both panels, the vertical white region indicates the previous estimates of the glass transition $\Gamma_g=195\pm 5$ \cite{PhysRevX.5.041033}. $k_g$ vanishes in a range of $\Gamma$ consistent with the onset of the glass transition. \textbf{(b)} Normalized instantaneous speed of collective shear waves $\bar{v}$ as a function of the inverse $\Gamma$ parameter, where the normalization is with respect to the maximum value of the instantaneous speed.}
    \label{fig:2}
\end{figure}

In the amorphous solid phase (panel (a), $\Gamma = 423$), shear modes exhibit a wave-like, propagating dispersion, consistent with predictions from elasticity theory and previously investigated in detail in Ref. \cite{PhysRevLett.109.178301}. In panel (b), at $\Gamma = 152$, slightly below the reported glass transition point $\Gamma_g \approx 195$, in the liquid phase, the dispersion relation undergoes a qualitative change at small wave vectors ($k a < 0.5$), where the emergence of a gap, consistent with Eq. \eqref{eq2}, becomes apparent.

This feature becomes even more pronounced in panel (c), deep in the liquid phase ($\Gamma = 64$), where the $k$-gap is clearly visible and larger than in panel (b). This trend supports the formation of a wave-vector gap in the dispersion of collective shear modes as the system transitions from the high-$\Gamma$ amorphous solid to the low-$\Gamma$ liquid. This transition is further corroborated by the behavior of the pair distribution function $g(r)$ shown in the inset of Fig.~\ref{fig1}(c), which reflects the structural changes across the glass transition.

After validating the emergence of a gap in the dispersion of collective shear modes with decreasing control parameter $\Gamma$, we extended our analysis to a broader range of $\Gamma$ values, spanning those shown in panels (a) to (c) of Fig.~\ref{fig1}. For each case, we extracted the corresponding dispersion relation and fitted the low-$k$ region using the $k$-gap equation, Eq. \eqref{eq2}. An example of this fitting procedure is presented in the inset of Fig. \ref{fig1}(b) for $\Gamma = 152$. We emphasize that due to the finite size of the experimental system, there exist lower bounds for both the wave vector and the frequency, denoted as $k_{\text{min}}$ and $\omega_{\text{min}}$, respectively. These cutoffs can be directly estimated and must be properly accounted for in the analysis (see more details in Appendix \ref{app1}). In the inset of Fig. \ref{fig1}(b), $k_{\text{min}}$ has been subtracted from the values on the $x$-axis, while $\omega_{\text{min}}$ is indicated by a dashed horizontal line. Notably, this line aligns well with the frequencies observed in the low-$k$ experimental data below the $k$-gap.

This analysis allowed us to determine the wave-vector gap $k_g$, the instantaneous shear velocity $v$ and the shear relaxation time $\tau$ as a function of $\Gamma$.

\begin{figure}
    \centering
    \includegraphics[width=\linewidth]{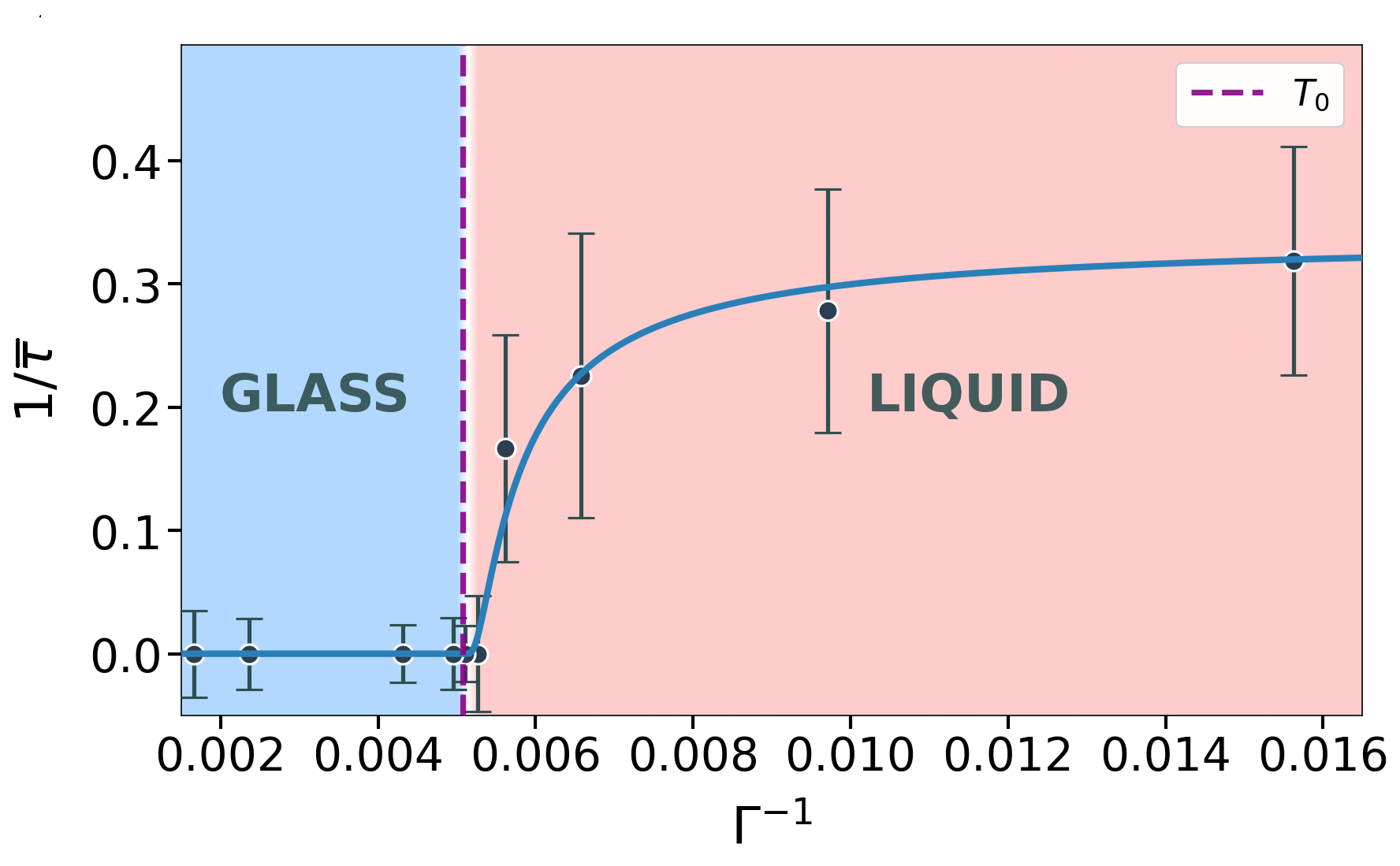}
    \caption{The extracted inverse relaxation time $1/\bar{\tau} \equiv 2 \bar{v} k_g$ as a function of the inverse Gamma parameter. The white band indicates the previous estimate of the glass transition $\Gamma_g=195\pm 5$ \cite{PhysRevX.5.041033}. The solid blue line is a fit of the experimental data to the VFT equation, Eq.~\eqref{vfteq}. The vertical purple dashed line is the fitted value of the Vogel temperature $T_0\equiv \Gamma^{-1}_0$, with $\Gamma_0 \approx200$.}
    \label{fig:3}
\end{figure}

In panel (a) of Fig.~\ref{fig:2}, we show the values of the wave-vector gap $k_g$ as a function of $\sim \Gamma^{-1}$, that is proportional to the temperature $T$ \eqref{defgamma}. We observe that $k_g$ diminishes by increasing $\Gamma$ towards the amorphous solid phase and drops rather quickly to approximately vanishing values around $\Gamma \approx 190$, a value that is consistent with the estimate glass transition temperature $\Gamma_g =195\pm 5$ \cite{PhysRevX.5.041033}. In the amorphous solid phase, $k_g$ vanishes up to the numerical accuracy and the finite size effects of our data. We note that the extracted values of the $k$-gap fall within the range $k_g a \in [0, 0.8]$ for the $\Gamma$ values explored. This range aligns remarkably well with values observed in other systems featuring vastly different interparticle interactions and particle sizes, suggesting a possible underlying universality, as originally proposed in \cite{PhysRevD.103.086001}.

By fitting the experimental data with Eq.~\eqref{eq2}, we extract the instantaneous shear velocity \(v\), which is directly related to the instantaneous shear modulus \(G_{\infty}\), as a function of \(\Gamma\). The results, shown in Fig.~\ref{fig:2}(b), reveal that \(v\) remains finite in both the amorphous solid and liquid phases but is significantly larger in the solid. Deep in the liquid state (small \(\Gamma\)) and deep in the solid state (large \(\Gamma\)), \(v\) varies only weakly with \(\Gamma\). In contrast, a sharp crossover appears near the glass transition, at \(\Gamma_g \approx 195 \pm 5\) \cite{PhysRevX.5.041033}, where \(v\) rapidly drops from its solid-like to liquid-like value.

Combining the extracted values of \(k_g\) and \(v\), we determine the shear relaxation time \(\tau = 1/(2 k_g v)\) as a function of \(\Gamma\). The experimental data for \(1/\tau\) versus \(\Gamma^{-1}\) are shown in Fig.~\ref{fig:3}. We find that \(1/\tau\) remains finite in the liquid phase but drops sharply as the glass transition is approached, signaling a rapid growth of the shear relaxation time. Below the glass transition, in the amorphous solid state, \(1/\tau\) is essentially zero within experimental accuracy. Assuming that \(\tau\) corresponds to the Maxwell time scale \(\sim \eta\), this behavior reflects the dramatic increase of the shear viscosity at and below the glass transition.

To describe this behavior theoretically, we use the well-established Vogel–Fulcher–Tammann (VFT) equation, which captures the super-Arrhenius temperature dependence of relaxation times in glass formers.  
Noting that \(\Gamma^{-1}\propto T\), we fit the experimental data for \(\tau\) with
\begin{equation}
\tau = A \exp\!\left(\frac{B}{\Gamma^{-1}-\Gamma^{-1}_0}\right),\label{vfteq}
\end{equation}
where \(A\), \(B\), and \(\Gamma_0\equiv 1/T_0\) are fitting parameters.  
Here \(A\) is the high-temperature limit of \(\tau\), \(B\) quantifies the effective energy barrier, and \(\Gamma^{-1}_0<\Gamma^{-1}_g\) represents the Vogel temperature at which \(\tau\) formally diverges, often identified with the ideal glass transition.

As shown in Fig.~\ref{fig:3}, the experimental data for $1/\tau$ are consistent with the super-Arrhenius behavior described by the VFT equation, Eq.~\eqref{vfteq}. In particular, our best fit gives:
\begin{equation}
    A=0.34,\qquad B=0.0006,\qquad \Gamma_0=200.
\end{equation}
This implies that the Vogel temperature is, as expected, slightly (around $3 \%$) below the glass transition temperature. 

This analysis reveals that the shear gap $k_g$ approaches zero continuously across the glass transition, consistent with the abrupt but continuous growth of the shear viscosity $\eta$ towards the glass transition. Moreover, this continuous trend is well captured by a super-Arrhenius increase of the relaxation time, consistent with the VFT equation characteristic of glass-forming materials.

\section{Outlook}
In summary, we have presented experimental evidence for the emergence of a wave-vector gap in the dispersion relation of collective shear modes across the glass transition in a two-dimensional mesoscopic colloidal system. In the liquid state, the form of the dispersion is consistent with predictions from Maxwell-Frenkel theory, and the onset of the $k$-gap coincides with the glass transition, as independently determined from the vanishing of the static shear modulus.

This shear gap vanishes continuously as the glass transition is approached, mirroring the sharp yet continuous growth of the shear viscosity near $T_g$. The experimental data are consistent with a super-Arrhenius temperature dependence of the shear relaxation time, inferred from the shear gap and well described by the Vogel–Fulcher–Tammann (VFT) equation. They further yield a rough estimate of the Vogel temperature, slightly below previously reported values for the onset of the glass transition.

Our findings not only provide direct experimental confirmation of existing theoretical predictions, but also suggest that this framework remains valid beyond conventional first-order solid-liquid melting, extending into continuous glass transition and very likely also 2D melting scenarios.

In the future, it would be interesting to independently measure the viscosity of our experimental system and verify directly the behavior of the shear relaxation time $\tau$ obtained indirectly in Fig.~\ref{fig:3}.

\section*{Acknowledgments}
JB and MB acknowledge the support of the Shanghai Municipal Science and Technology Major Project (Grant No.2019SHZDZX01). MB acknowledges the sponsorship from the Yangyang Development Fund, P.K. from DFG, project 453041792 (Heisenberg funding).

%
\appendix
\section{On the minimal frequency and wave-vector}
\label{app1}
\begin{figure}[ht]
    \centering
    \includegraphics[width=\linewidth]{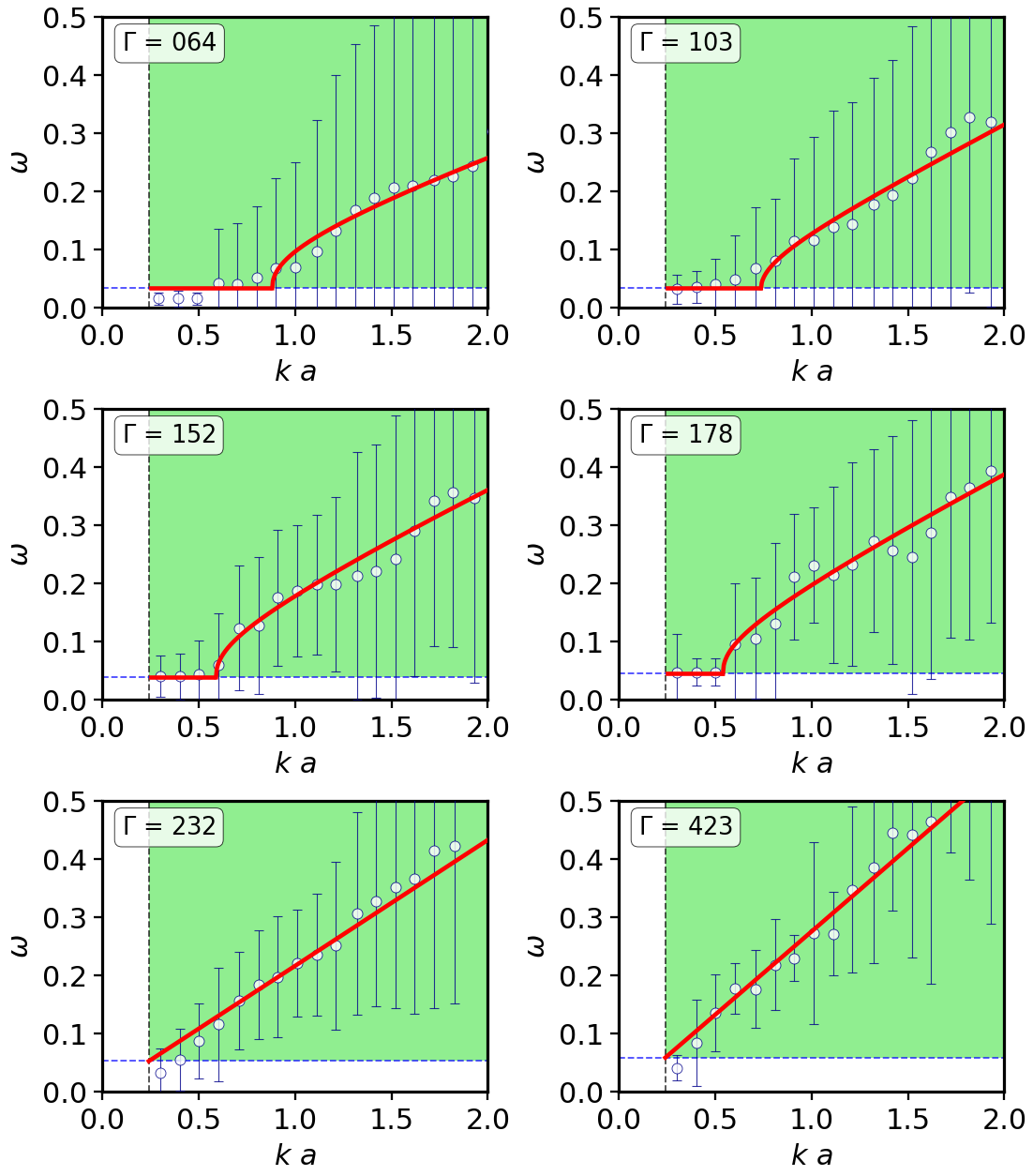}
    \caption{White symbols denote the transverse dispersion relation extracted from the experimental data, together with the corresponding error bars. The vertical and horizontal dashed lines indicate the cutoff values $\omega_{\mathrm{min}}$ and $k_{\mathrm{min}}$, respectively, as determined in Appendix~\ref{app1}. The red lines show fits to the gapped dispersion relation given in Eq.~\eqref{eq2}, performed exclusively within the green shaded region.
}
    \label{fig:4}
\end{figure}
Due to finite-size effects, the minimum wave vector accessible in our analysis is
\begin{equation}
    k_{\mathrm{min}} = 2\pi \sqrt{\frac{1}{L_x^2} + \frac{1}{L_y^2}},
\end{equation}
where $L_x$ and $L_y$ denote the linear dimensions of the $L_x \times L_y$ observation window used in the experimental analysis.

The corresponding minimum frequency is related to this wave vector through $\omega_{\mathrm{min}} = v k_{\mathrm{min}}$, where $v$ is the speed of sound in the system. The effects of these finite-size cutoffs, $k_{\mathrm{min}}$ and $\omega_{\mathrm{min}}$, are illustrated in Fig.~\ref{fig:4}, where they are indicated by dashed lines. For sufficiently large values of $\Gamma$ in the solid phase, the experimental data are consistent with a linear dispersion relation of the form
\[
\omega - \omega_{\mathrm{min}} = v \left( k - k_{\mathrm{min}} \right),
\]
confirming the validity of our estimates for these two scales.

In the main text, the analysis of the shear-excitation dispersion is therefore restricted to the range $\omega > \omega_{\mathrm{min}}$ and $k > k_{\mathrm{min}}$, highlighted by the green shaded region in Fig.~\ref{fig:4}.

\end{document}